\documentclass[aps,prl,twocolumn,longbibliography,superscriptaddress,floatfix]{revtex4-1}
\usepackage{amsfonts}
\usepackage{mathtools}
\usepackage{graphicx}
\usepackage{epsfig}
\usepackage{dcolumn}
\usepackage{bm}
\usepackage{amsmath}
\usepackage{graphicx}
\usepackage[latin1]{inputenc}
\usepackage{ulem}
\usepackage{epstopdf}
\usepackage{subfigure}
\usepackage{color}
\usepackage{amsthm}
\usepackage{newlfont}
\usepackage{graphicx}
\usepackage{epstopdf}
\usepackage{appendix}
\usepackage[breaklinks=true]{hyperref}
\usepackage{breakcites}
\usepackage{textcomp}
\usepackage{appendix}
\usepackage{multirow}	
\usepackage{color}
\usepackage{amssymb}
\usepackage{epsfig}
\usepackage{bm}
\usepackage[american]{babel}
\usepackage{braket}
\hypersetup{colorlinks=true,linkcolor=blue,citecolor=blue,filecolor=blue,urlcolor=blue,pdfstartview=FitH}
	
\begin{document}
\title{
Weakly nonergodic dynamics in the Gross--Pitaevskii lattice
}
\author{Thudiyangal Mithun}
    \altaffiliation{These authors contributed equally to this work}
 \affiliation{Center for Theoretical Physics of Complex Systems, Institute for Basic Science, Daejeon 34051, Korea}   
\author{Yagmur Kati}%
    \altaffiliation{These authors contributed equally to this work}
    \affiliation{Center for Theoretical Physics of Complex Systems, Institute for Basic Science, Daejeon 34051, Korea}
\affiliation{Basic Science Program, Korea University of Science and Technology (UST), Daejeon 34113, Republic of Korea}
\author{Carlo Danieli}%
\affiliation{Center for Theoretical Physics of Complex Systems, Institute for Basic Science, Daejeon 34051, Korea}
\author{Sergej Flach}%
\affiliation{Center for Theoretical Physics of Complex Systems, Institute for Basic Science, Daejeon 34051, Korea}
\begin{abstract}
The microcanonical Gross--Pitaevskii (aka semiclassical Bose-Hubbard)  lattice model dynamics is characterized by a pair of energy and norm densities. The grand canonical Gibbs distribution fails to describe a part of the density space,
due to the boundedness of its kinetic energy spectrum. We define Poincare equilibrium manifolds and compute the statistics of microcanonical excursion times off them.
The tails of the distribution functions quantify the proximity of the many-body dynamics to a weakly-nonergodic phase, which occurs when the average excursion time is infinite. We find that a crossover to weakly-nonergodic dynamics takes place {\sl inside} the nonGibbs phase, being {\sl unnoticed} by the largest Lyapunov exponent.
In the ergodic part of the non-Gibbs phase, the Gibbs distribution should be replaced by an
unknown modified one.
We relate our findings to the corresponding integrable limit, close to which the actions are interacting through
a short range coupling network.
\end{abstract}
\maketitle
\pagestyle{myheadings}

Equipartition and thermalization are cornerstone concepts of understanding stability and predictability of complex matter dynamics. Proximity to integrable limits may have a strong impact on the needed time scales, or even on equipartition itself. Let us consider a dynamical system which is characterized by a countable set of preserved actions at the very integrable limit,
as e.g. for harmonic lattice vibrations in crystals.
Close to the limit, nonintegrable couplings between the actions induce a nontrivial dynamics of the latter. The nonintegrable couplings define a certain connectivity network on the action lattice.

The nonlinear coupling network of the actions can be
{\sl long ranged}. That is precisely the case with translationally invariant weakly nonlinear lattice wave equations,
or phonon dynamics in crystals, or
e.g. the celebrated Fermi-Pasta-Ulam (FPU) chain \cite{Fermi:1955,Ford:1992}. Then the linear integrable limit yields actions which are related to standing or plane waves (harmonic phonons) which traverse the entire system. Weak local nonlinearities therefore induce a coupling network
which is long ranged \cite{Ford:1992}.
At whatever small, but finite, energy densities in an equipartitioned state, all plane waves and thus actions will be coupled
regardless of their characteristics (e.g. the eigenfrequency). Selection rules due to momentum conservation
do not alter the above argument.
Nature nicely confirms that, since phonon dynamics
in crystals appears to be equipartitioned down to the smallest temperatures. At the same time, approaching zero densities
will lead to a diminishing of the largest Lyapunov exponent, and thus equipartition times are expected to smoothly diverge
in the very limit.

The focus of this work is the case of a Gross--Pitaevskii (GP), aka Bose-Hubbard (BH), lattice with local nonlinear many-body interactions, and short range
hoppings. In the limit of {\sl large} densities the nonlinear interactions dominate over the hoppings, the actions turn local in real space, and the system disintegrates
into an uncoupled set of strongly anharmonic oscillators in real space. Close to the limit the short range hoppings induce
a nonintegrable {\sl short range} coupling network between the actions.
Anomalous and potentially nonergodic large density dynamics was reported for the GP lattice  \cite{Rasmussen:2000,Rumpf:2004,Rumpf:2008,Rumpf:2009},
including nonequilibrium transport properties \cite{Iubini:2012,Iubini:2013} and self localization \cite{Livi:2006, Holger:2013, Kruse:2017}.
Indications for nonergodic dynamics were also observed for similar model classes \cite{Ivanchenko:2004,Pino:2016}. 

Strict nonergodic dynamics implies a separation of the phase space into disjoint parts under the action of
Hamiltonian dynamics, which could imply the presence of additional symmetries. Such symmetries are unlikely to
be restored upon the smooth change of control parameters. An alternative scenario is observed in glassy dynamics,
as e.g. shown by Bouchaud via the appearance of consecutive metastable states, whose lifetimes are distributed according to power-law distributions \cite{Bouchaud:1992}. If the average lifetime of the metastable states turns infinite,
a trajectory might still visit almost all the phase space, however strictly infinitely long time is required to observe that when
computing averages. Such dynamics, while
formally being ergodic, turns {\it nonergodic} for any finite averaging time. 
Similar behavior has been discussed by Eli, Rebenshtok and Barkai in a set of papers dedicated to continuous-time random walks \cite{bel2005weak,bel2006random,rebenshtok2007distribution}. Therein, the phenomenon goes under the name of {\it weak ergodicity breaking}, or {\it weak nonergodicity}.
Lutz further formalized the connection between power-law distribution  and weak nonergodicity in the context of optical lattices \cite{lutz2004power}.


The goal of this work is to show the existence of a weak nonergodic phase of
the GP lattice dynamics  and to quantitatively
assess the crossover line from an ergodic to   a weak nonergodic regime in the relevant two-dimensional density parameter space. The GP lattice dynamics is conserving energy and norm (particle number). The microcanonical
dynamics is depending on the corresponding pair of densities. If the dynamics is ergodic, the time average of an observable (a function of the phase space coordinates) should exist and be equal to a phase space average with
a proper distribution function. Assuming equal weight of microstates, the Boltzmann (canonical) or Gibbs
(grand canonical) distributions are the proper choice. Rasmussen et al. showed that the Gibbs distribution
with positive temperature and arbitrary chemical potential is addressing only a part of the accessible
microcanonical density space
\cite{Rasmussen:2000}. Negative temperatures yield divergent partition functions, and a proper nonGibbs distribution for the complementary space
is not known. In that nonGibbs density space the microcanonical dynamics is characterized by anomalous fluctuations, slow relaxations, and potentially (weakly) nonergodic dynamics. We note that the mere fact of a nonGibbs
regime is not sufficient to conclude that the dynamics is nonergodic, since the analysis is based solely on phase space
integrations and does not consider any aspect of the accompanying dynamics.

Our strategy is to use proper observables $f$ as functions of the phase space variables. Assuming ergodicity we
may obtain the expected phase space average $\bar{f}$. The condition $f=\bar{f}$ defines an {\sl equilibrium
Poincare manifold} of co-dimension 1 which separates the accessible microcanonical phase space into two
disjoint sets. By assumption of ergodicity, a microcanonical trajectory must pierce this manifold infinitely
many times during its evolution, to ensure that the microcanonical time average $\langle f \rangle = \bar{f}$.
Let us consider the event of two consecutive piercings, and the trajectory excursion off the manifold in between.
We will assess the statistics, correlations, and other properties of these excursions. At variance with correlation function
computations, our strategy allows to return to individual excursions which contribute to a particular feature.
In a recent study \cite{Danieli:2017} of a finite FPU system, an entropy function on the system phase space
was used as an observable $f$. This integral quantity becomes insensitive to relevant nonergodic fluctuations in the
limit of large volume $N$. The key ingredient in this work is to use simultaneously all observables which correspond to
integrals of motion in the large density limit. The piercings of one single trajectory through $N$ equilibrium manifolds
will then be analyzed.

\begin{figure}
\includegraphics[width=0.99 \columnwidth, height=0.25 \textheight]{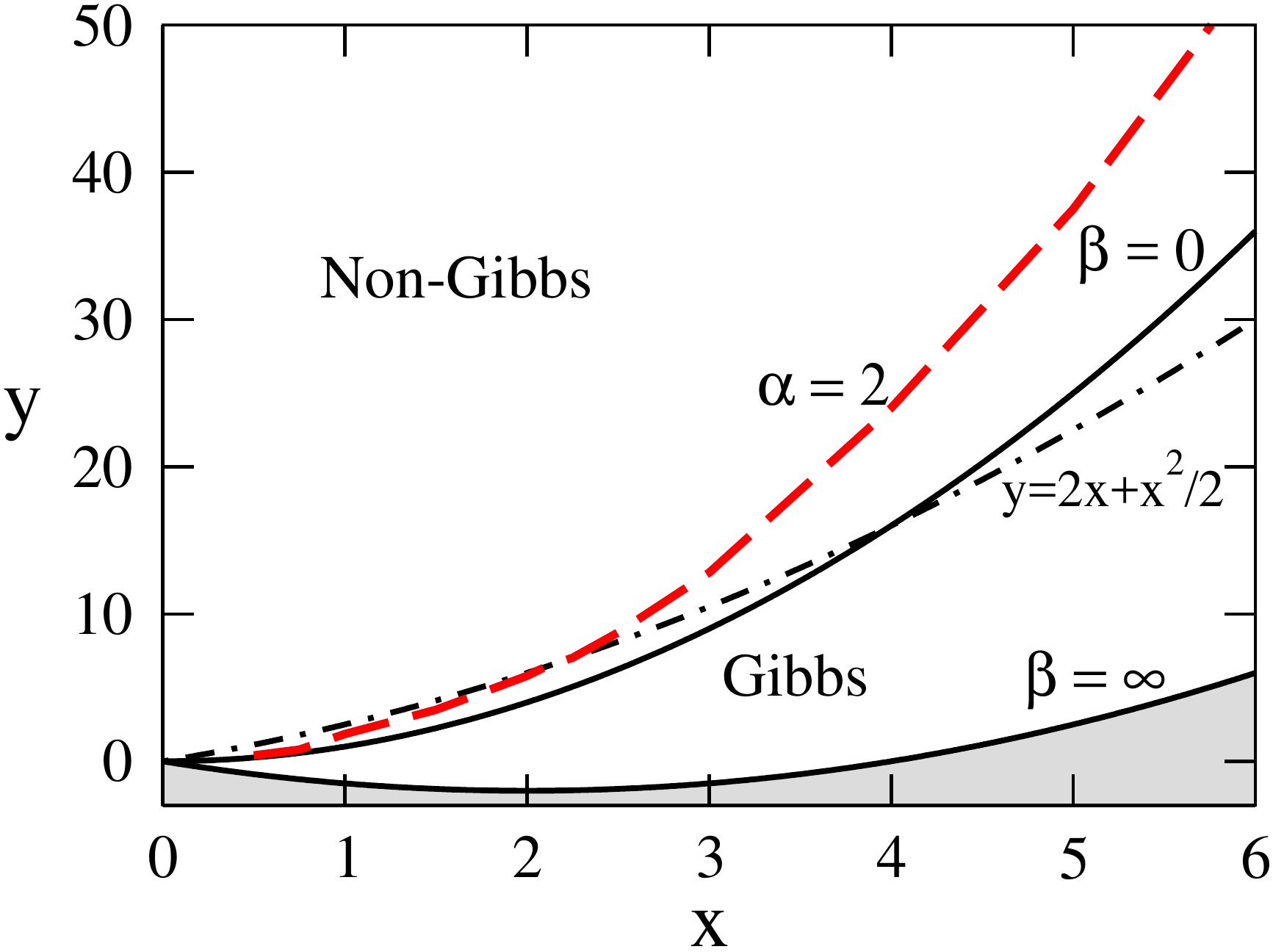}
\caption{\label{fig:dnls1}(Color online) GP phase diagram in the microcanonical density space $(x,y)$. Thick solid lines $y_{GS}=-2x+x^2/2$ and $y_{nG}=x^2$  are for
$\beta=\infty$ ($T= 0$) and $\beta=0$ ($T=\infty$), respectively. No microcanonical states exist below the $\beta=\infty$ line and this area is shaded.
Microcanonical states above the $\beta=0$ line exist, but are not addressable through a Gibbs distribution.
Dashed-dotted line $y_h=2x+x^2/2$, above which microcanonical states with constant norm density
$|\psi_m | = const$ cease to exist. Thick dashed line (red): ergodic to nonergodic transition dynamics where $\alpha=2$ (see text for details).
}
\end{figure}

The one-dimensional GP lattice equations read
\begin{equation}
 i\frac{\partial \psi_m}{\partial t}+(\psi_{m+1}+\psi_{m-1})-g|\psi_m|^2\psi_m=0,
\label{eq:DNLS1}
\end{equation}
where $m$ labels the lattice sites, and $g$ is a nonlinear parameter related to the two-body scattering length.
Eq.(\ref{eq:DNLS1}) is generated by the Hamiltonian equations of motion  $i\dot{\psi}_m=\frac{\partial \mathcal{H}}{\partial \psi_m^{\ast}}$ with the Hamiltonian
\begin{equation}
\mathcal{H} =\sum_{m}\big[-(\psi_m^{\ast}\psi_{m+1}+\psi_m \psi_{m+1}^{\ast})+\frac{g}{2}|\psi_m|^4 \big].
\label{eq:DNLS2}
\end{equation}
Here $\psi_m^{\ast}$ and $\psi_m$ are pairs of conjugated phase space variables, the sum runs over $N$ lattice sites, and periodic boundary conditions $\psi_1 =\psi_{N+1}$ are used.
In addition to the total energy $\mathcal{H}$,
the above equations also conserve
the norm $\mathcal{A}=\sum_m |\psi_m|^2$ which is the classical analogue to the quantum mechanical total number of particles. The canonical transformation $\psi_m=\sqrt{A_m}\exp(i\phi_m)$ maps Eq.(\ref{eq:DNLS2}) into
\begin{equation}
\mathcal{H} =\sum_{m}\big[-2\sqrt{A_m A_{m+1}}\cos(\phi_m-\phi_{m+1})+\frac{g}{2}|A_m|^2 \big].
\label{eq:DNLS_rotated1}
\end{equation}
Rasmussen et al. \cite{Rasmussen:2000} used Eq.(\ref{eq:DNLS_rotated1}) to compute the classical grand-canonical partition function
\begin{equation}
\textit{Z}= \int_{0}^{\infty} \int_{0}^{2\pi} \prod_{m=1}^N d\phi_m dA_m \exp^{-\beta (\mathcal{H}+\mu \mathcal{A})}.
\label{eq:Part_funct_Z}
\end{equation}
Here $\mu$ is the chemical potential and $\beta$ the inverse temperature $\beta = 1/T \geq 0$.
The mapping of the pair of Gibbs parameters $\{\beta,\mu\}$ onto the microcanonical density space
$\{h,a\}$ with $h=\mathcal{H}/N$ and $a=\mathcal{A}/N$ leaves a part of the high energy density space unaddressed,
with the infinite temperature $\beta=0, \beta \mu=const$ line being the border between the addressable density space part (Gibbs regime) and the complementary one (nonGibbs regime)  \cite{Rasmussen:2000}.
It is convenient to use rescaled densities $x = g a$, $y = g h$. Then the Gibbs part of the density space is sandwiched between the zero temperature $\beta=\infty$ line $y_{GS}=-2x+x^2/2$ and the infinite temperature $\beta=0$ one
$y_{nG}=x^2$ in Fig.\ref{fig:dnls1}.
It was conjectured that microcanonical dynamics in the nonGibbs phase is
nonergodic due to the observed formation of concentrated hot spots of localized norm/energy excitations
\cite{Rasmussen:2000}. These excitations appear to be related to exact discrete breather solutions
\cite{Flach:1998,Campbell:2004,Flach:2008,Franzosi:2011}.
Interestingly these exact finite energy time-periodic solutions are continuable into single site anharmonic oscillator
excitations in the integrable limit of infinite densities, coined anticontinuous limit by MacKay and Aubry \cite{Mackay:1994}. Rumpf developed an entropic picture of fragmentation of the field into two components
in the nonGibbs regime - a condensate of the above hot spots, and a remaining thermalized
component with infinite temperature $\beta=0$ \cite{Rumpf:2004,Rumpf:2008}.
Whether the spots thermalize and whether the system is ergodic or not, remained unaddressed.
This leads to the question, whether the GP lattice turns nonergodic precisely in the nonGibbs
regime. Below we will study lifetime distributions of the hot spots, show that these times
stay finite inside a part of the nonGibbs regime, and discuss the consequences.
We also note that homogeneous norm density states $\psi_m = \sqrt{a} e^{i \phi_m}$
with $\phi_m=0$ minimize the energy at a given value of $x$ and yield the $\beta=\infty$ line.
At the same time, the largest energy of these states is obtained for $\phi_m=\pi$ and yields
the line $y_h=2x+x^2/2$, which is located in the nonGibbs regime for $x\leq 4$ in Fig.\ref{fig:dnls1}.
Thus nonGibbs dynamics can be generated with initial states which are completely homogeneous
in their norm and energy density distributions. For energy densities $y > y_h$ no homogeneous
states are available.

In all simulations shown in this work, Eq.(\ref{eq:DNLS1}) is integrated by using the symplectic procedure $SBAB_2$ described in Ref.\cite{Skokos:2009}, with time step $\Delta t =0.02$, which keeps the relative energy error below $0.1\%$ (the total norm is conserved up to computational roundoff precision). The observables are simply the local norm densities $f_n=g|\psi_n|^2$, $n=1,...,N$ which turn
into integrals of motion in the infinite density limit.
They define $N$ ergodic Poincar\'e sections $\mathcal{F}_n : f_n=x$.
Unless specified otherwise, we consider $N = 2^{10}$ sites.
We integrate a trajectory and
track the times $t_i^{(n)}$ the trajectory pierces any of the equilibrium manifolds $\mathcal{F}_n$.
The \textit{excursion times} follow as $\tau^{(n,\pm)}(i)=t^{(n)}_{i+1}-t_i^{(n)}$
where the sign $\pm$ is set by the sign of  $(f_n-x)$ during the excursion and tells whether we
monitor an excursion with local augmentation (+) or depletion (-) of the norm density.
We then obtain the probability distribution functions (PDF) of the excursion times $P_{\pm}(\tau)$.
We attempt to fit the PDF tails with a power law $P(\tau) \propto \tau^{-\alpha}$ to find the
dependence of $\alpha$ on the densities $(x,y)$. For $\alpha \leq 2$ we  conclude that the dynamics
is weakly nonergodic, since the average of the excursion times $\langle \tau \rangle$ diverges. 

We note that we can not exclude the presence of exponential cutoffs
in the unresolvable part of $P$ at large values of $\tau$. We checked that the precise form of the chosen initial states is not relevant in the ergodic regime. All that matters are the values of $x$ and $y$.
We further compute the maximal Lyapunov Characteristic Exponent (mLCE) - the average rate of divergence of nearby trajectories, which is a quantitative measure of the degree of nonintegrability and deterministic chaos \cite{Skokos:2016}.
We numerically solve the tangent dynamics of a small amplitude perturbation $\chi_m(t)$ to a given
(numerically obtained) trajectory $\{\psi_m(t)\}$ \cite{Skokos:2016}  by integrating
\begin{equation}
i\dot{\chi}_m=-(\chi_{m+1}+\chi_{m-1})+g (2|\psi_m|^2\chi_m+\psi_m^2\chi_m^{\ast}).
\label{eq:DNLSlce}
\end{equation}
The mLCE follows as $\Lambda(t)=\lim_{t\to\infty} \frac{1}{t}\ln\frac{||\chi(t)||}{||\chi(0)||},$ where $||\chi(t)|| =\sqrt{\sum_{m=1}^N|\chi_m(t)|^2}$ (see e.g. \cite{Johansson:2010}).
Details of the integration scheme are given in the supplemental material \cite{Supple}.
In practice, we need finite but large enough
averaging times on which the $\Lambda(t)$ saturates \cite{Supple}.

\begin{figure}[h]
\includegraphics[width=0.99 \columnwidth]{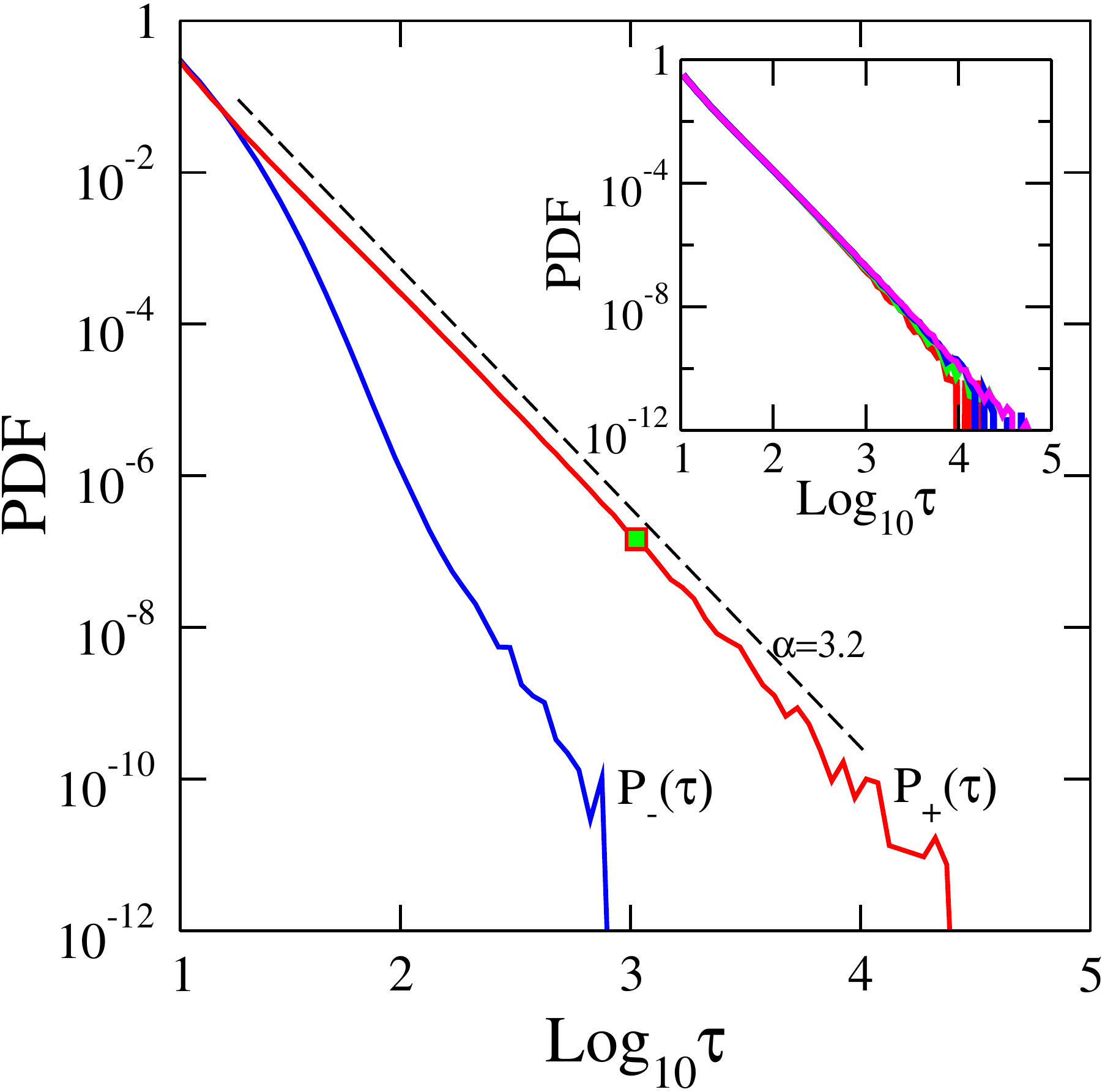}
\\
\includegraphics[width=0.99 \columnwidth]{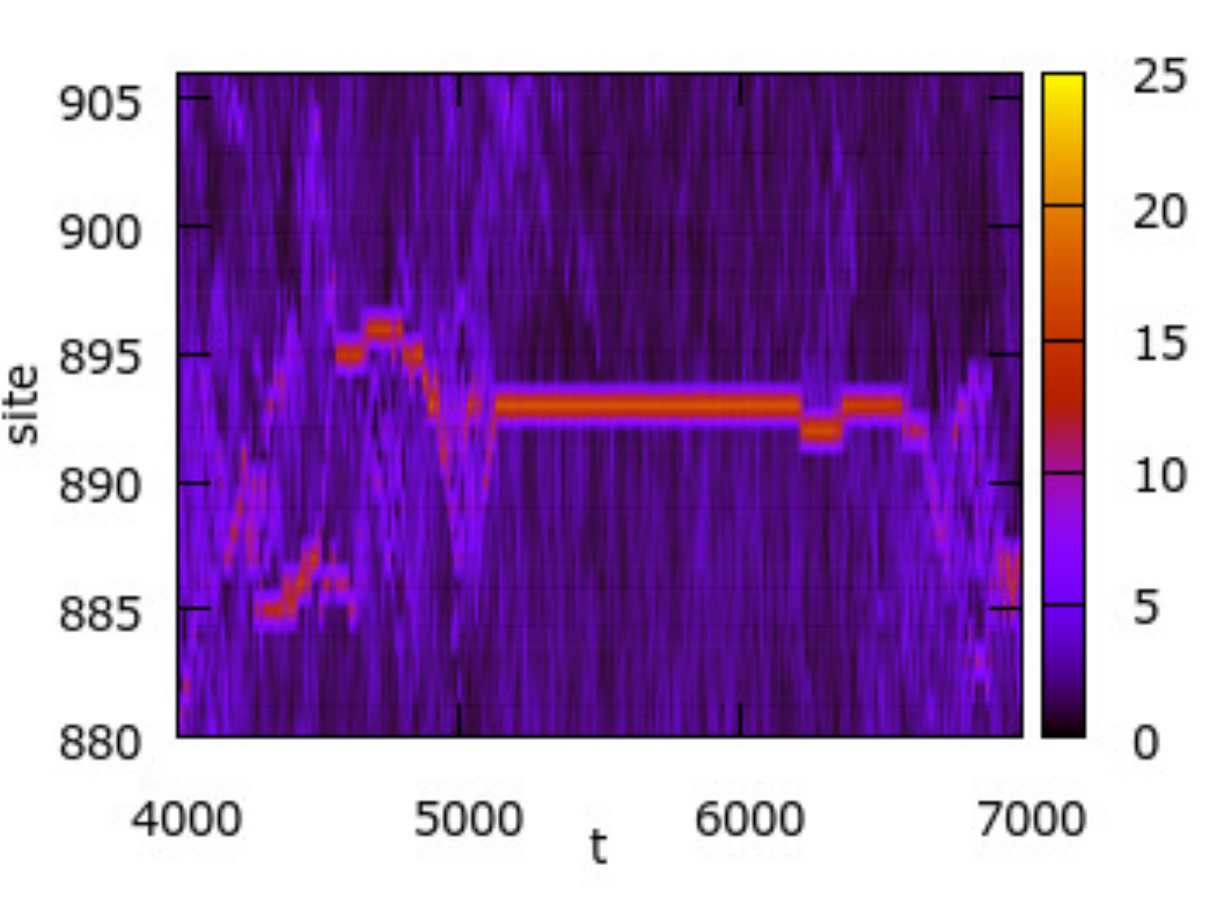}
 \caption{\label{fig:dnls2}
(Color online) (a) $P_{\pm}(\tau)$ for $x=2,~y=4$ (on the $\beta=0$ line) for $N=2^{10}$. The upper red curve indicates $P_{+}(\tau)$ and bottom blue
curve for $P_{-}(\tau)$. The dashed line is an algebraic decay $\tau^{-\alpha}$ with  $\alpha=3.2$. Inset: $P_{+}(\tau)$ for different system sizes $N = 512,1024,2048,4096$).
 (b) Time evolution of density, $|\psi_n|^2$ in correspondence of one of the excursion time $\tau$ marked with the green colored square in Fig. \ref{fig:dnls2}(a). Here $g=1$.
}
\end{figure}

In Fig.\ref{fig:dnls2}(a) we show $P_{\pm}(\tau)$ for a density pair $x=2,~y=4$ on the $\beta=0$ line
$y_{nG}(x)$. We observe that $P_{+}(\tau)$ (upper red curve) has a clear algebraic tail, while
$P_{-}(\tau)$ (lower blue curve) decays much faster, and in a more complex manner.
In the following we will present results for the exponent $\alpha$ for $P_+(\tau)$ only,
which reads $\alpha = 3.2 \pm 0.1$.
By our definition, the dynamics is ergodic, despite being on the border line to the nonGibbs
phase. We plot in the inset of Fig.\ref{fig:dnls2}(a) the function
$P_+(\tau)$ obtained for different volumes $N=512,1024,2048,4096$ and conclude that we can exclude
the impact of finite size effects.
In Fig.\ref{fig:dnls2}(b), we show the time evolution of the norm density of one of the excursions
which contribute to the algebraic tail (marked with the green square in Fig.\ref{fig:dnls2}(a)).
We observe the generation of a long lasting discrete breather like excitation out of ergodic fluctuations,
which persists for a large time $10^3$ and finally decays again into the thermalized surrounding.
We positively tested this conclusion for many other tail excursions.



\begin{figure}
\includegraphics[width=0.99 \columnwidth]{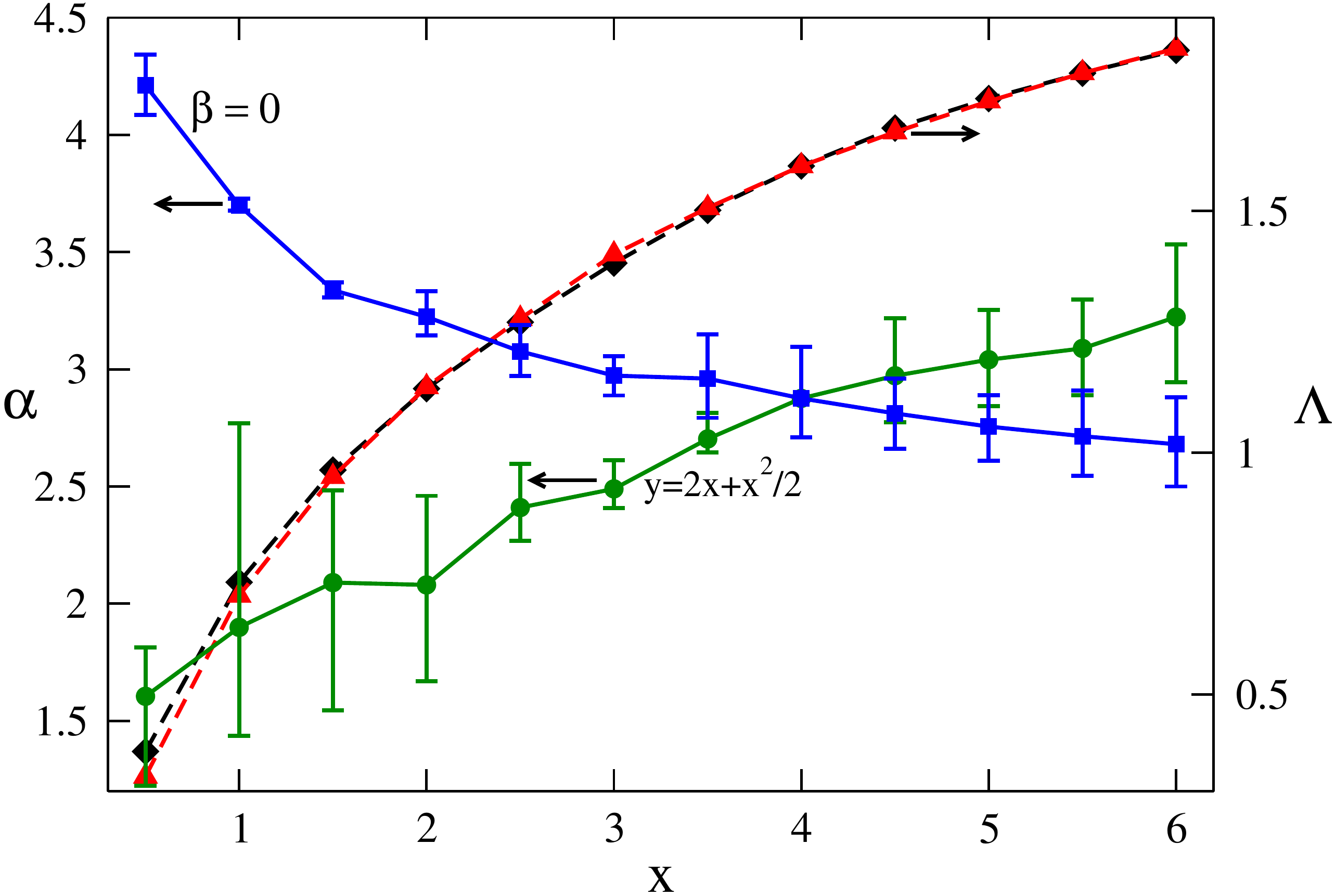}
\caption{\label{fig:dnls4}{ (Color online) The exponent $\alpha$ of the power-law tail for the norm density $x$ calculated along the
two lines $y_{nG}=x^2$ (blue solid line) and $y_h=2x+\frac{x^2}{2}$ (green solid line). The dashed black and red lines represent the mLCE ($\Lambda(x)$)
along the lines $y_{nG}$ and $y_h$, respectively.
}}
\end{figure}

In Fig.\ref{fig:dnls4} we present results for the exponent $\alpha$ along the
two characteristic lines $y_h(x)$ and $y_{nG}(x)$. The function $\alpha_{nG}(x)$ along the
$\beta=0$ line $y_{nG}(x)=x^2$ (which separates Gibbs and nonGibbs phases)
monotonously decreases with {\sl increasing} $x$. Its value is clearly $\alpha > 2$ in the whole
assessed range
$0 < x < 6$. We may anticipate that weak nonergodicity ($\alpha=2$) happens around $x\sim 20-30$
in that line. The function $\alpha_h(x)$ along the limiting line for homogeneous states $y_h(x)=
2x+x^2/2$ is monotonously decreasing with {\sl decreasing} $x$. While increasing density $x$ 
enhances ergodicity on that line, we observe a transition to weak nonergodicity $\alpha_h < 2$ for
$x < 2$. Let us remind that in the weakly nonergodic regime every observable becomes trajectory dependent for any finite averaging time. This dependence translates into large uncertainties in their measurement, which herewith results in the large error bars in Fig.\ref{fig:dnls4} for $\alpha\sim 2$ (details of the estimate of the exponent $\alpha$ are given in the supplemental material \cite{Supple}).
The transition to weak nonergodicity happens well in the nonGibbs phase. 
Therefore we conclude that parts of the nonGibbs phase allow for ergodic dynamics. 
This in turn implies that a new nonGibbs distribution function should exist. We also mapped the density space points which correspond to $\alpha=2$ into a line $y_{NE}(x)$ in Fig.\ref{fig:dnls1}. This line is clearly located {\sl inside}
the nonGibbs phase, albeit close to its boundary.

The Lyapunov exponent function $\Lambda_{nG}(x)$ along the $\beta=0$ line $y_{nG}(x)$,
and the function $\Lambda_h(x)$ along the $y_h(x)$ line, are plotted in Fig.\ref{fig:dnls4}
and show no anomalies, neither in the ergodic, nor in the weakly nonergodic, neither in the Gibbs,
nor in the nonGibbs phases. Therefore, we conclude that weakly nonergodic dynamics is triggered
by local fluctuations (discrete breather like excitations) which leave a part of the system
well thermalized in between them. The Lyapunov exponent is sensitive to the chaotic
dynamics in these thermalized puddles, but is not sensitive to the presence of weakly nonergodic
boundaries between the puddles.

Let us discuss our observations. The Gross--Pitaevskii lattice model is one of the remarkable cases where the large density limit yields an integrable system of disconnected
anharmonic oscillators. The network of nonintegrable perturbations which couple the actions off that limit is given by the hopping part of the GP model, and is short ranged.
As a consequence, the microcanonical dynamics becomes weakly nonergodic at large but finite densities for a macroscopic system, which is still at a finite distance
from some integrable limit. We quantify these observations by computing distributions of excursion times off
equilibrium Poincare manifolds, and measuring the exponents in their tails. Long excursion times are related to the generation of hot spots, or discrete breather like excitations.
Our method is therefore able to quantitatively assess discrete breather lifetimes at equilibrium.
Note that weakly nonergodic dynamics is going well along with nonzero Lyapunov exponents. This happens because a part of the system condenses into discrete breather like regions, or
spots of regular dynamics, while regions between these spots still evolve in a chaotic fashion.
%

It is tempting to relate these observations to the existence of a nonGibbs phase in the microcanonical GP lattice dynamics. Note that this nonGibbs phase existence follows from
the existence of a second conserved quantity (the norm aka particle number) and is a result of a purely statistical analysis.
We find that a part of the nonGibbs phase is
ergodic. Therefore we conclude that a yet unknown new grand canonical distribution function might exist which describes the equilibrium and ergodic dynamics there. Nevertheless, the explicit form of this distribution function is not known. At the same time,
we expect that weakly nonergodic dynamics due to large densities will also take place in the Gibbs part of the microcanonical control parameter space.

The microcanonical thermodynamical description as well as the existence of negative temperature for Hamiltonian systems with bounded spectrum have been questioned by Duenkel \textit{et. al.} in \cite{Dunkel:2014}. Their argument says that, in order to describe the thermodynamics of such systems, the Gibbs entropy has to be employed, which implies the non-existence of negative temperature.
In our case, we followed the microcanonical ergodic dynamics defined with the Boltzmann temperature and is proven to be valid in the defined Gibbs phase. Moreover, it has been recently shown that the Boltzmann entropy (which admits negative temperatures) provides the correct description of the microcanonical thermodynamics of systems like the GP \cite{Buonsante:2016,Buonsante:2017} (further discussions can be found in \cite{Rumpf:2004,Rumpf:2008,Rumpf:2009,Iubini:2012,Iubini:2013}).

To conclude, we applied a novel method of statistical analysis of excursion times off equilibrium Poincare manifolds to the transition from ergodic to nonergodic dynamics in the Gross--Pitaevskii lattice model. Our results are in analogy with the weak nonergodicity phenomena studied in glass systems \cite{Bouchaud:1992}, continuous-time random walks \cite{bel2005weak,bel2006random,rebenshtok2007distribution}, as well as in other many-body systems \cite{Danieli:2017}. We expect them to be applicable also to larger spatial dimensions, and to other lattice models with similar integrable limits.
We also speculate that spatial disorder, which induces Anderson localization, at small densities will again lead to weakly nonergodic dynamics at (then small but) finite densities.


\section{Acknowledgements}
The authors acknowledge financial support from IBS (Project Code No. IBS-R024-D1). We thank I. Vakulchyk, A. Andreanov and C. H. Skokos for helpful discussions.

\clearpage
\newpage
\begin{center}
\textbf{\large Supplemental Material for: 'Weakly nonergodic dynamics in the Gross--Pitaevskii lattice'}
\end{center}

\section{LCE Calculation}\label{ap2}
The variational equation 
\begin{equation}
i\dot{\chi}_m=-(\chi_{m+1}+\chi_{m-1})+g (2|\psi_m|^2\chi_m+\psi_m^2\chi_m^{\ast}),
\label{eq:DNLSlce}
\end{equation}
 is solved by using symplectic $SBAB_2$ integrator scheme. The Hamiltonian,
  \begin{equation}
 \mathcal{H} =\sum_{m}\big[-(\chi_m^{\ast}\chi_{m+1}+\chi_m \chi_{m+1}^{\ast})+\frac{g}{2}(4|\psi_m|^2|\chi_m|^2+\psi_m^2\chi_m^{\ast 2} )\big]
\label{eq:b1}
\end{equation}
corresponds to Eq. (\ref{eq:DNLSlce}) is split as
  \begin{eqnarray}
    A =\sum_{m}\big[-(\chi_m^{\ast}\chi_{m+1}+\chi_m \chi_{m+1}^{\ast})\big],\\
   B =\frac{g}{2}\sum_{m}\big[(4|\psi_m|^2|\chi_m|^2+\psi_m^2\chi_m^{\ast 2})\big].
  \label{eq:b2}
  \end{eqnarray}
B can be written as $B=P+Q$, where
\begin{equation}
\small P=2g \sum_{m}|\psi_m|^2|\chi_m|^2,~~Q=\frac{g}{2}\sum_{m}\psi_m^2\chi_m^{\ast 2}
\label{eq:b3}
\end{equation}
\subsubsection{Action of the operator $e^{\tau L_A}$ on $\chi_m$}
Fast Fourier transform (FFT) is used for Hamiltonian $A$.
  \begin{eqnarray}
    \chi_q=\sum_{l=1}^N \chi_l e^{2 \pi i (q-1) (l-1)/N},\\
   \chi_q^{'}=\chi_q e^{2 i \cos(2\pi(q-1)/N)\tau},\\
   \chi_m^{'}=\frac{1}{N}\sum_{q=1}^N \chi_q^{'} e^{-2 \pi i (m-1) (q-1)/N},
  \label{eq:b4}
  \end{eqnarray}
    where $\chi_m^{'}$ is $\chi_m$ at $t+\tau$.
\subsubsection{Action of the operator $e^{\tau L_P}$ on $\chi_m$}
 It can be solved exactly as
\begin{equation}
\chi_m^{'}=\chi_me^{-2gi |\psi_m|^2\tau}
\label{eq:b5}
\end{equation}
 \subsubsection{Action of the operator $e^{\tau L_Q}$ on $\chi_m$}
     \begin{equation}
\chi_m^{'}=(\chi_{am}+i \chi_{bm}),
\label{eq:b6}
\end{equation}
where
 \begin{eqnarray}
  \chi_{am}= c_1 \cosh(A \tau)+\frac{1}{A}\sinh(A\tau)\big[bc_1-a c_2\big],\\
  \chi_{bm}= c_2 \cosh(A\tau)-\frac{1}{A}\sinh(A\tau)\big[bc_2+a c_1\big],
\label{eq:b7}
 \end{eqnarray}
  where $A=\sqrt{a^2+b^2}$, $a=Re(g \psi_m^2)$, $b=Im(g \psi_m^2)$, $c_1=Re(\chi_m)$ and $c_2=Im(\chi_m)$.  Equations are integrated by using the $SBAB_2$ scheme given in \cite{Bodyfelt:2011}.

\begin{figure}[pht]
   \includegraphics[width=8.8cm,height=7.0cm]{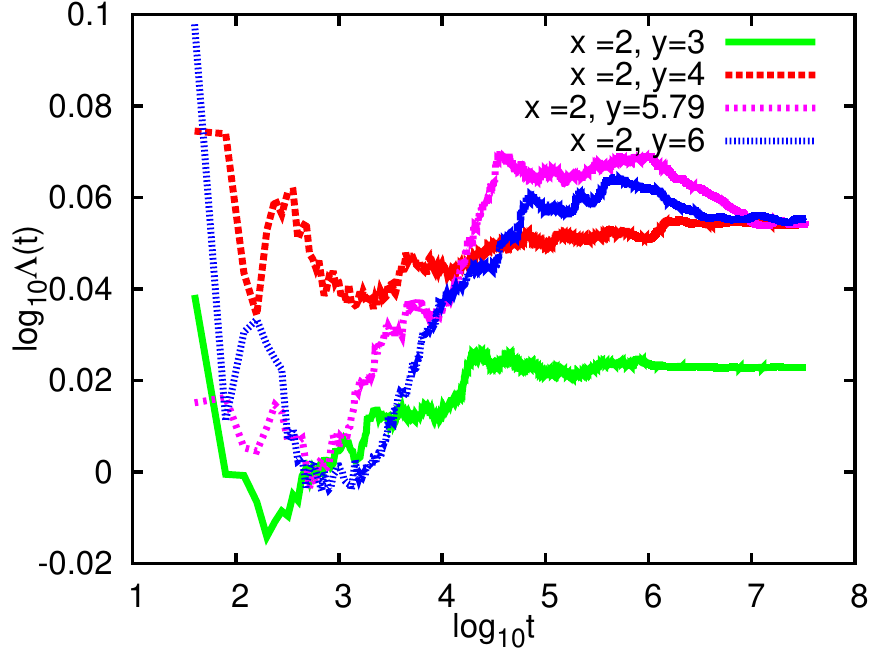}
 \caption{\label{fig:dnls6}{\footnotesize (Color online) Time evolution of the maximal Lyapunov exponent for the fixed norm density $x=2$. The different y values represent the Gibbs regime ($y=3$), the phase transition line ($y=x^2$), the ergodic to nonergodic transition line ($y = 5.79$) and the nonGibbs regime.}}
\end{figure}

 \begin{center}
    \begin{table}[h]
    \caption{\label{tab:dnls2} Calculation of $\Lambda(t)$}
    \begin{tabular}{p{0.8cm} | p{1.2cm} p{1.2cm}| p{2.0cm} c }
    \hline
    \hline
   x    & $y=x^2$ & $\Lambda$ &$y=2x+\frac{x^2}{2}$ & $\Lambda$   \\
        &         &            &        &              \\ [0.5ex] \hline
   0.5  &   0.25  &  0.3823    &1.125    &   0.3308         \\
   1.0  &   1.0   &  0.7323    &2.5      &   0.7049        \\
   1.5  &   2.25  &  0.9642    &4.125    &   0.9505        \\
   2.0  &   4.0   &  1.1328    &6        &   1.1369       \\
   2.5  &   6.25  &  1.2701    &8.125    &   1.2774       \\
   3.0  &   9.0   &  1.3922    &10.5     &   1.4107        \\
   3.5  &   12.25 &  1.5017    &13.125   &   1.5066        \\
   4.0  &   16.0  &  1.5932    &16       &   1.5932        \\
   \hline    \hline
    \end{tabular}
    \end{table}
    \end{center}
In Fig. (\ref{fig:dnls6}), the evolution of mLCE is depicted for norm density $x=2$. The energy density varies from Gibbs regime to nonGibbs regime. The mLCE saturates after $t=10^7$ and we calculate it by taking the average of mLCE for the range $t=10^7-10^8$. The positive value of mLCE says that the system is chaotic in both Gibbs regime and nonGibbs regime. The comparison of mLCE at Gibbs regime, at $y=4$ and  $y=3$, reveals that the chaos in the system is directly proportional to the energy density and such dependency is observed for the problem of quartic Klein-Gordon chain of coupled anharmonic oscillators \cite{Skokos:2013}. Tab. \ref{tab:dnls2} shows the mLCE for the various values of norm density. We could see that chaos is proportional to the norm density too. The main result is that mLCE in the nonergodic regime approaches its value at the phase transition line. It elucidates that change from the ergodic to nonergodic transition is extremely slow.

\section*{Fitting $\alpha$ in the tails of $P_+(\tau)$}

\begin{figure}[pht]
   \includegraphics[width= 0.95\columnwidth]{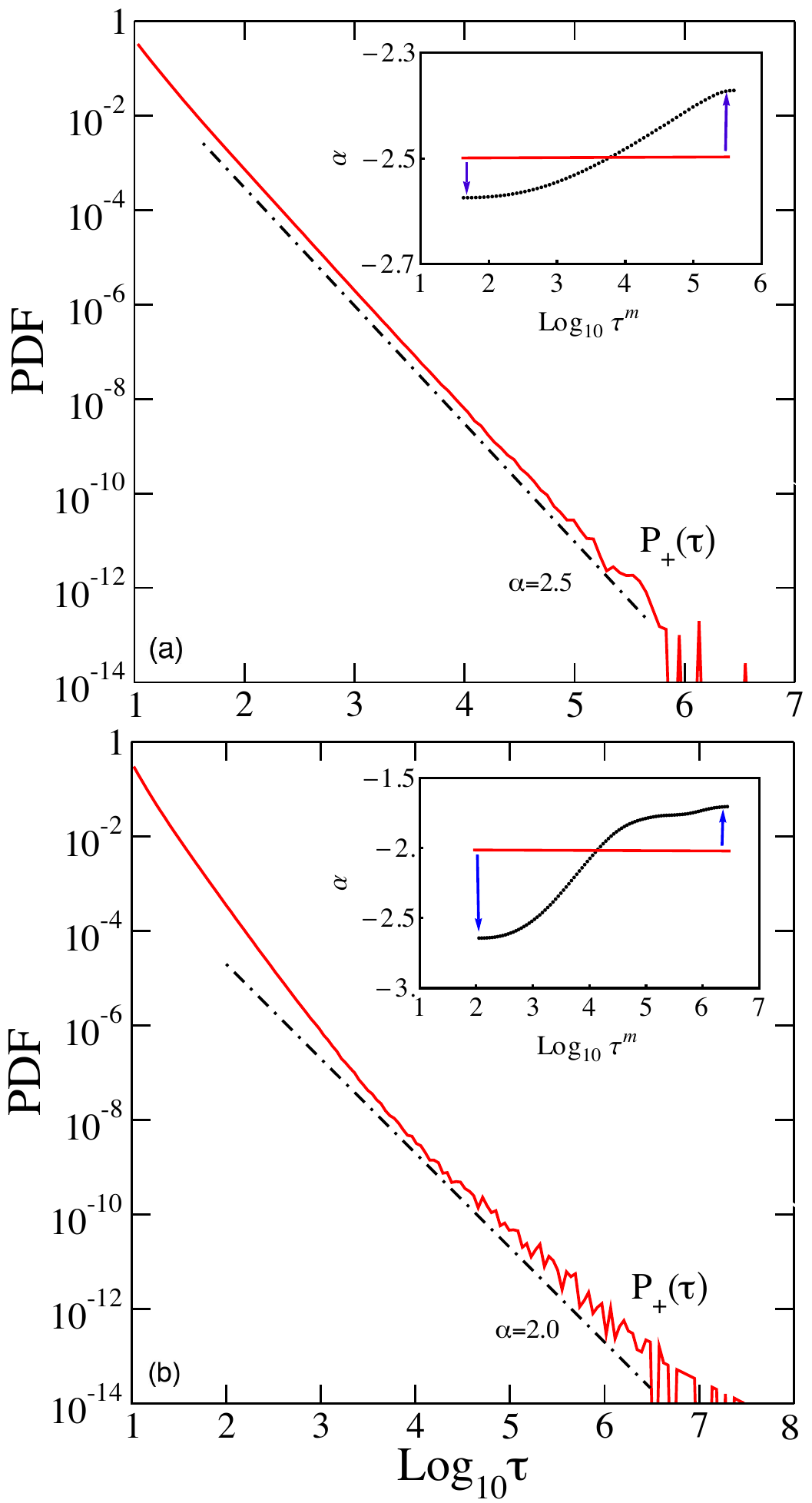}
 \caption{\label{fig:dnls7}
 {\footnotesize (Color online) PDF $P_+(\tau)$ obtained for $(x,y) = (3,10)$ and $(x,y) = (1.5,4.125)$. Insets: exponent $\alpha$ versus the lower bound $\tau^m$.
 }}
\end{figure}

The PDF $P_+(\tau)$ is obtained using bins equispaced on a logarithmic scale. 
We estimate the exponent $\alpha$ by smoothening the curve $P_+(\tau)$ and using the Hodrick-Prescott filter method \cite{Hodrick:1997}. The 
extrapolation method has been used on intervals in the tail of the PDF, by fixing the upper interval bound $\tau^{M}$ and varying the lower 
bound $\tau^m$. The upper bound $\tau^M$ is defined by the first bin which counts zero events. 
In Fig.\ref{fig:dnls7} we show two PDFs which correspond to two cases on the green line of Fig.3 of the main text, for $(x,y) = (3,10)$ (Fig.\ref{fig:dnls7}(a)) and $(x,y) = (1.5,4.125)$ (Fig.\ref{fig:dnls7}(b)). 
In the insets, we show the exponent $\alpha$ measured as a function of the lower bound $\tau^m$. The horizontal line represents  the final measured 
value, while the arrows represent the largest distances from the final value. These distances then become the error bars in Fig.3 of the main text. 
In Fig.\ref{fig:dnls7}(a), the system is ergodic ($\alpha = 2.5$) and the PDF shows a power-law trend.
In Fig.\ref{fig:dnls7}(b), the PDF shows a more complex trend. 
The curve is actually nowhere really close to a power law, and could decay even slower than any algebraic decay.
The above defined error becomes of the order of the mean, which in fact means that the measured exponent is
not very meaningful. 

\bibliographystyle{apsrev4}
\let\itshape\upshape
\normalem
\bibliography{reference1}
\end{document}